\documentclass[preprint,showpacs,preprintnumbers,amsmath,amssymb,nofootinbib]{revtex4}
\usepackage{amssymb}
\usepackage{graphicx}
\usepackage[mathscr]{eucal}

\def\be{\begin{equation}}
\def\ee{\end{equation}}
\def\ba{\begin{eqnarray}}
\def\ea{\end{eqnarray}}

\newcommand\scri{\mathscr{I}}

\begin{document}

\preprint{IGPG04/8-4, AEI--2004--072}
\title{Black hole evaporation: A paradigm}
\author{Abhay\ Ashtekar${}^{1}$ and Martin Bojowald${}^{2,1}$}
\affiliation{1. Institute for Gravitational Physics and Geometry,\\
Physics Department, Penn State, University Park, PA 16802, USA\\
2. Max-Planck-Institut f\"ur Gravitationsphysik,
Albert-Einstein-Institut, Am M\"uhlenberg 1, D-14476 Potsdam,
Germany}

\begin{abstract}

A paradigm describing black hole evaporation in non-perturbative
quantum gravity is developed by combining two sets of detailed
results: i) resolution of the Schwarzschild singularity using
quantum geometry methods \cite{m,ab1}; and ii) time-evolution of
black holes in the trapping and dynamical horizon frameworks
\cite{sh1,sh2,ak1,ak2}. Quantum geometry effects introduce a major
modification in the traditional space-time diagram of black hole
evaporation, providing a possible mechanism for recovery of
information that is classically lost in the process of black hole
formation. The paradigm is developed directly in the Lorentzian
regime and necessary conditions for its viability are discussed.
If these conditions are met, much of the tension between
expectations based on space-time geometry and structure of quantum
theory would be resolved.
\end{abstract}

\pacs{0460P, 0470D}

\maketitle

\section{Introduction}
\label{s1}

In classical general relativity, a rich variety of initial data on
past null infinity, ${\scri^-}$, can lead to the formation of
a black hole.%
\footnote{For simplicity of discussion, in this article we will
consider only zero rest mass matter fields and assume that past
null infinity is a good initial value surface. To include massive
fields, one can suitably modify our discussion by adjoining past
(future) time-like infinity to past (future) null infinity.}
Once it is formed, space-time develops a new, future boundary at
the singularity, whence one can not reconstruct the geometry and
matter fields by evolving the data \emph{backward} from future
null infinity, ${\scri^+}$. Thus, whereas an appropriately chosen
family of observers near ${\scri}^-$ has full information needed
to construct the entire space-time, no family of observers near
${\scri^+}$ has such complete information. In this sense, the
classical theory of black hole formation leads to information
loss. Note that, contrary to the heuristics often invoked (see,
e.g. \cite{swh2}), this phenomenon is not directly related to
black hole uniqueness results: it occurs even when uniqueness
theorems fail, as with `hairy' black holes \cite{hairy} or in
presence of matter rings non-trivially distorting the horizon
\cite{gh}. The essential ingredient is the future singularity,
hidden from ${\scri^+}$, which can act as the sink of information
(see, in particular, Penrose's remarks in \cite{rp1}.)

\begin{figure}
\begin{center}
\includegraphics[height=2in]{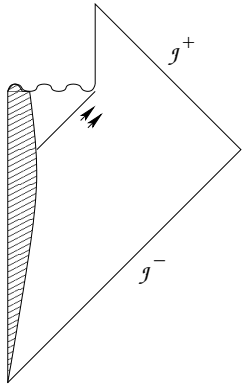}
\caption{The standard space-time diagram depicting black hole
formation and evaporation.} \label{Traditional}
\end{center}
\end{figure}

A natural question then is: what happens in quantum gravity? Is
there again a similar information loss? Hawking's celebrated work
of 1974 \cite{swh1} analyzed this issue in the framework of
quantum field theory in curved space-times. In this approximation,
three main assumptions are made: i) the gravitational field can be
treated classically;  ii) one can neglect the back-reaction of the
spontaneously created matter on the space-time geometry; and iii)
the matter quantum field under investigation is distinct from the
collapsing matter, so one can focus just on spontaneous
emission.%
\footnote{Generally, only the first two assumptions are
emphasized. However, we will see that the third also has a bearing
on the validity of semi-classical considerations.}
Under these assumptions, Hawking found that there is a steady
emission of particles to ${\scri^+}$ and the spectrum is thermal
at a temperature dictated by the surface gravity of the final
black hole. In particular, pure states on ${\scri^-}$ evolve to
mixed states on ${\scri^+}$. In a next step, one can include
back-reaction. To our knowledge, a detailed, systematic
calculation is still not available. In essence one argues that, as
long as the black hole is large compared to the Planck scale, the
quasi-stationary approximation should be valid. Then, by appealing
to energy conservation and the known relation between the mass and
the horizon area of \emph{stationary} black holes, one concludes
that the area of the event horizon should steadily decrease. This
then leads to black hole evaporation depicted in figure
\ref{Traditional} \cite{swh1}.

If one does not examine space-time geometry but uses instead
intuition derived from Minkowskian physics, one may be surprised
that although there is no black hole at the end, the initial pure
state has evolved in to a mixed state. Note however that while
space-time is now dynamical even after the collapse, \emph{there
is still a final singularity, i.e., a final boundary in addition
to} $\scri^+$. Therefore, it is not at all surprising that, in
this approximation, information is lost ---it is still swallowed
by the final singularity \cite{rp1}. Thus, provided figure
\ref{Traditional} is a reasonable approximation of black hole
evaporation and one does not add new input `by hand', then pure
states must evolve in to mixed states.

The question then is to what extent this diagram is a good
representation of the physical situation. The general argument in
the relativity community has been the following (see e.g.
\cite{rmw1}). Figure \ref{Traditional} should be an excellent
representation of the actual physical situation as long as the
black hole is much larger than the Planck scale. Therefore,
problems, if any, are associated \emph{only} with the end point of
the evaporation process. It is only here that the semi-classical
approximation fails and one needs full quantum gravity. Whatever
these `end effects' are, they deal only with the Planck scale
objects and would be too small to recover the correlations that
have been steadily lost as the large black hole evaporated down to
the Planck scale. Hence pure states must evolve to mixed states
and information is lost.

Tight as this argument seems, it overlooks two important
considerations. First, one would hope that quantum theory is free
of infinities whence figure \ref{Traditional} can not be a good
depiction of physics near the \emph{entire singularity} ---not
just near the end point of the evaporation process. Second, the
event horizon is a highly global and teleological construct. (For
a recent discussion of limitations of this notion, see
\cite{ak3}). Since the structure of the \emph{quantum} space-time
could be very different from that of figure \ref{Traditional} near
(and `beyond') the singularity, the causal relations implied by
the presence of the event horizon of figure \ref{Traditional} is
likely to be quite misleading. Indeed, Hajicek \cite{ph} has
provided explicit examples to demonstrate that the Vaidya
solutions which are often used to model the evaporating black hole
of figure \ref{Traditional} can be altered just in a Planck scale
neighborhood of the singularity to change the structure of the
event horizon dramatically and even make it disappear.

The purpose of this article is to point out that these
considerations are important and conclusions drawn from figure
\ref{Traditional} are therefore incomplete. More precisely, we
will argue that \emph{the loss of information is not inevitable
even in space-time descriptions favored by relativists.} As in
other discussions of the black hole evaporation process, we will
not be able to present rigorous derivations. Rather, we will
present a paradigm%
\footnote{In this article, the term `paradigm' is used in the
modest, dictionary sense, `One that serves as a pattern or model'.
The paradigm presented here was briefly sketched in section 8 of
\cite{ak3}.}
by drawing on two frameworks where detailed and systematic
calculations have been performed: i) analysis of the fate of the
Schwarzschild singularity in loop quantum gravity; and ii) the
dynamical horizon formalism which describes evolving black holes
in classical general relativity. Even without these details,
certain general conclusions could be drawn simply by assuming that
the space-time geometry is somehow modified near the singularity
and analyzing the Hawking process on this new space-time. But then
there is a multitude of possibilities. As we will see below, loop
quantum gravity and dynamical horizon considerations serve to
focus the discussion and suggest concrete directions for future
work. The manner in which black hole (and cosmological)
singularities are resolved in loop quantum gravity provides a
specific type of quantum extension of space-time and the fact that
thermodynamical considerations apply also to dynamical horizons
makes it plausible to think of the Hawking process as evaporation
of these quasi-local horizons. The final result of these
considerations is a space-time description of black hole
evaporation in the physical, Lorentzian setting in which one
allows for a \emph{quantum} extension of the space-time geometry
\emph{beyond} singularity. Since the space-time no longer has a
future boundary at the singularity, pure quantum states on
${\scri}^-$ can evolve to pure quantum states on ${\scri}^+$.

The plausibility of this scenario is supported by the fact that
its 2-dimensional version is realized \cite{av} in the CGHS black
hole \cite{cghs}. (For earlier work along these lines, see
especially \cite{am}.) There, it is possible to isolate the true
degree of freedom and carry out an exact quantization using, e.g.,
Hamiltonian methods. On the resulting Hilbert space, one can in
particular define the quantum (inverse) metric operator. The
classical black hole metric arises as the expectation value in a
suitable quantum state, i.e., in the \emph{mean field
approximation.} Hawking effect emerges through the study of small
fluctuations on this mean field. One can explicitly check that
this mean field approximation is good in a significant portion of
the quantum space-time. However, the quantum fluctuations are very
large near the entire singularity, whence the approximation fails
there. The quantum (inverse) metric operator itself is
well-defined everywhere; only its expectation value vanishes at
the classical singularity. Thus, quantum geometry is defined on a
manifold which is \emph{larger} than the black hole space-time of
the mean field approximation. The mean field metric is
well-defined again in the asymptotic region `beyond' the
singularity.%
\footnote{There is a qualitative similarity with the theory of
ferromagnetism. The (inverse) metric is analogous to the
magnetization vector. If you have a large ferromagnet (such as the
earth) a small, central portion of which is heated beyond the
Curie temperature, the mean field approximation will hold far away
from this central region and the magnetization operator will have
a well-defined mean value there. That region is analogous to the
part of the full, quantum space-time where there is a well-defined
classical metric. The analysis of the Hawking effect is analogous
to that of spin-waves on this part of the ferromagent, where the
mean field approximation holds. While the mean field approximation
fails in the central region where the expectation value of
magnetization vanishes,  quantum theory provides a good
description of the entire magnet, including the central region, in
terms of microscopic spins.}
Thus, there is a single asymptotic region in the distant past
\emph{and} distant future and pure states on ${\scri}^-$ evolve to
pure states on $\scri^+$ of the full quantum space-time.

In this paper, we will focus on 4 dimensions where the qualitative
picture is similar but the arguments are based on a number of
assumptions. We will spell these out at various steps in the
discussion. As we will see, specific calculations need to be
performed to test if the assumptions are valid and the scenario is
viable also in 4 dimensions. Our hope is that the proposed
paradigm will provide direction and impetus for the necessary
detailed analysis which will deepen our understanding of the
evaporation process, irrespective of whether or not the paradigm
is realized.

The paper is organized as follows. In section 2, we summarize the
resolution of the Schwarzschild singularity by effects associated
with the quantum nature of geometry. The new paradigm for black
hole evaporation is presented in section 3. Section 4 contains
some concluding remarks.

\section{Quantum geometry and the Schwarzschild interior}
\label{s2}

Since the key issues involve the final black hole singularity and
since this singularity is expected to be generically space-like
(see, e.g. \cite{md}), the situation is similar to cosmology. In
fact, the interior of the Schwarzschild horizon is naturally
foliated by 3-manifolds which are spatially homogeneous with the
Kantowski-Sachs isometry group. Accordingly, the result of absence
of singularities in homogeneous loop quantum cosmology \cite{b3}
can be applied to this situation of the Kantowski-Sachs
`mini-superspace' of vacuum, spatially homogeneous space-times.
That this is possible has been shown explicitly using
Arnowitt-Deser-Misner (ADM) variables \cite{m}: unlike classical
evolution, the dynamical quantum equation does not break down at
the location of the classical singularity. However, since the ADM
variables allow only non-degenerate metrics, the geometrical
meaning of the resulting space-time extension has remained obscure
in this framework. The connection dynamics phase space, by
contrast, is an extension of the ADM phase space where the
(density weighted) triad, which captures the Riemannian geometry,
is allowed to vanish. Thanks to this larger phase space, the
extended space-time has a clearer interpretation: the `other side'
of the singularity corresponds to the new domain of the enlarged
phase space where the triad reverses its orientation.%
\footnote{In addition, the elementary variables that feature in
the quantization used in \cite{m} ---the exponentials of $i$ times
extrinsic curvature components--- do not have natural analogs in
full geometrodynamics based on the ADM variables. In the
connection mini-superspace, by contrast, the elementary variables
are just holonomies of homogeneous connections, i.e., restrictions
to the basic variables used in the full theory to the symmetry
reduction under consideration.}
Therefore, in this section we will summarize the results that have
been obtained in the connection-dynamics framework \cite{ab1}.

The first result is that, although the co-triad and curvature
diverge at the singularity in the classical theory, the
corresponding quantum operators are in fact bounded on the full
kinematic Hilbert space. This analysis is analogous to that which
established the boundedness of the quantum operator representing
the inverse scale factor in the spatially homogeneous, isotropic
quantum cosmology \cite{b1,abl}. As in that analysis, the co-triad
operator has various nice properties one expects of it and
departures from the classical behavior appear only in the deep
Planck regime (i.e. very near what was classical singularity).
This finiteness results from the fact that the `polymer
representation' of the Weyl relations underlying our quantum
description is inequivalent to the `standard representation' used
in quantum geometrodynamics (for details, see, e.g.,
\cite{alrev}). It is analogous to the fact that matter
Hamiltonians in the full theory are densely defined \cite{tV}
operators. This result suggests that quantum dynamics may well be
singularity-free. But a definitive conclusion can only be drawn
through a detailed analysis.

Using quantum geometry, one can write down a well-defined
Hamiltonian constraint. In the mini-superspace under
consideration, there are only two degrees of freedom. One can be
interpreted as the radius of any (round) 2-sphere in the slice and
the other (the norm of the translational Killing field) is a
measure of the anisotropy. It is natural to use the first as an
intrinsic `clock' and analyze how anisotropy `evolves' with
passage of this `time'. In quantum theory, one can expand out the
state $|\Psi\rangle$ as $|\Psi\rangle = \sum_{\phi,\tau}
\psi(\phi,\tau)|\phi, \tau\rangle$ where $\phi$ are eigenvalues of
the anisotropy operator and $\tau$ of the radius operator. The
Hamiltonian constraint is of the form:
\be \label{1} f_+(\tau)\, {\hat{O}_+}\, \psi(\phi, {\tau+2\delta})
+ f_o(\tau)\, {\hat{O}_o}\,
 \psi(\phi, {\tau}) + f_-(\tau)\,
{\hat{O}_-}\, \psi(\phi, {\tau-2\delta})  =0 \ee
where $f_{\pm}, f_o$ are rather simple functions of $\tau$,
$\hat{O}_\pm, \hat{O}_o$ are rather simple operators on functions
of $\phi$ alone and $\delta$ is a number whose value is determined
by the smallest area eigenvalue in Planck units. Being a
constraint, it simply restricts the physically allowed states.
However, one can also regard it as providing `time-evolution' of
the quantum state through discrete time steps of magnitude
$2\delta$ (in Planck units). The functions $f$ and the operators
$\hat{O}$ are such that this evolution does not break down at
$\tau =0$ (which corresponds to the classical singularity). Thus,
as in quantum cosmology \cite{b2,abl} one finds that the
\emph{quantum} evolution does not stop at the singularity; one can
evolve right through it \cite{ab1}. The state remains pure.
However one expects that, in the \emph{deep} Planck regime around
the singularity, the notion of a classical space-time geometry
would fail to make even an approximate sense in general.
Nonetheless, there is no longer a final boundary in the interior,
whence the full quantum evolution is quite different from the
classical one.

This calculation was done \cite{ab1} in the Kantowski-Sachs
mini-superspace and $|\Psi\rangle$ represents the state of the
Schwarzschild black hole interior in loop quantum gravity. This
black hole can not evaporate: there is no matter and, because of
the restriction to spherical symmetry, there can not be Hawking
radiation of gravitons either. However, since the generic
singularity is expected to be space-like (see, e.g., \cite{md}),
one may hope that the general intuition about the resolution of
the Schwarzschild singularity provided by this calculation can be
taken over to models in which gravity is coupled to scalar fields,
where the evaporation does occur. Indeed, there is already some
work on the spherical model without restriction to the interior
\cite{bs,hw,b4} and its extension is now in progress. The initial
results support expectations from the homogeneous models. Here, we
will assume that the \emph{overall, qualitative} features of our
singularity resolution will continue to be valid in these models.

\section{Evaporation Process}
\label{s3}

The physical situation we wish to analyze is the following: some
radiation field on ${\scri^-}$ collapses and forms a large,
macroscopic black hole which then evaporates. For simplicity, we
will restrict ourselves to the \emph{spherically symmetric sector
of Einstein gravity coupled to a massless Klein-Gordon field}. The
incoming state on ${\scri}^-$ will be assumed to be a coherent
state peaked at a classical scalar field representing a large
`pulse', i.e., a field which is large over a compact region of
${\scri}^-$ and vanishes (or become negligible) outside this
region. Note that there is a single scalar field, coupled to
gravity, whose collapse from ${\scri}^-$ leads to the formation of
the black hole and whose quanta are radiated to ${\scri}^+$ during
the evaporation process. There are no test fields; the system is
`closed'.

In this setting, conclusions drawn from classical general
relativity should be valid to an excellent approximation until we
are in the Planck regime near the singularity. Thus, marginally
trapped surfaces would emerge and their area would first grow. In
this phase the world tube of marginally trapped surfaces  would be
a \emph{trapping horizon} \cite{sh1}. For the massless scalar
field under consideration, during and for a long time after the
collapse, it would be space-like \cite{gv,acg} and thus constitute
a \emph{dynamical horizon} \cite{ak1,ak2}. When Hawking radiation
starts to dominate the in-falling scalar field, the trapping
horizon would be time-like and thus constitute a \emph{time-like
membrane} \cite{ak3}. In the spherical symmetric case now under
consideration, this scenario was discussed already in the eighties
(see, in particular \cite{jy,ph}). However, constructions were
tailored just to spherical symmetry and made use of some heuristic
considerations involving an `ergo-region of an approximate Killing
field.' Therefore, although well-motivated, the discussions
remained heuristic. Laws governing the growth of the area of
dynamical horizons and shrinkage of area of time-like membranes
are now available in a general and mathematically precise setting
\cite{ak2,sh2}. Furthermore, laws of black hole mechanics have
been extended to these dynamical situations. These results
strengthen the older arguments considerably and reenforce the idea
that what evaporates is the trapping horizon.

Let us now combine this semi-classical picture with the discussion
of section \ref{s2} on the resolution of the singularity to draw
qualitative conclusions on what the black hole evaporation process
would look like in full loop quantum gravity. Once the Planck
regime is reached, a priori there are two possibilities:\\
$\bullet\,\,$ a) States which start out semi-classical on
${\scri}^-$ never become semi-classical on the `other side' of the
singularity (say, in the sense discussed in \cite{bd,aps}).  Then
only a part of the process can be described in space-time terms.
However, one \emph{can} look at the problem quantum mechanically
and conclude that pure states remain pure. If we restricted them
only to the classical part of the space-time and measure
observables which refer only to this part, we would get a density
matrix but this is not surprising; it happens even in laboratory
physics when one ignores a part of the system. \\
$\bullet\,\,$ b) As in spatially homogeneous, isotropic
cosmologies coupled to a massless scalar field \cite{aps}, after
evolving through the deep Planck regime, the state becomes
semi-classical again on the `other side' so we can use a classical
space-time description also in `distant future'.

In the CGHS model, possibility b) holds. Furthermore, using the
underlying conformal structure, one can show that the classical
region in the distant future remains causally connected to that in
the distant past in the full quantum theory; there is no baby
universe. Such a calculation is yet to be undertaken in four
space-time dimensions. If it turns out that the possibility a)
holds, it would be impossible to speak of a scattering matrix
since there would not be an adequate ${\scri}^+$ or a space-like
surface in the distant future for the `final' states to live on.
Hence, it would be quite difficult to say anything beyond the
statement that pure states remain pure. If b) holds, one can
compare various scenarios. Therefore, in the rest of the article,
we will focus on b).

\begin{figure}
\begin{center}
\includegraphics[height=3in]{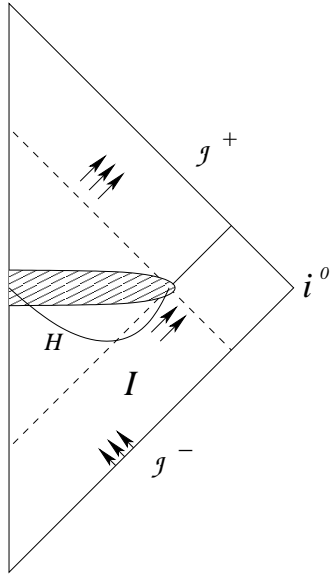}
\caption{Space-time diagram of black hole evaporation where the
classical singularity is resolved by quantum geometry effects. The
shaded region lies in the `deep Planck regime' where geometry is
genuinely quantum mechanical. $H$ is the trapping horizon which
is first space-like (i.e., a dynamical horizon) and grows because
of infalling matter and then becomes time-like (i.e., a time-like
membrane) and shrinks because of Hawking evaporation. In region I,
there is a well-defined semi-classical geometry.} \label{NonSingH}
\end{center}
\end{figure}

A space-time diagram that could result in scenario b) is depicted
in figure \ref{NonSingH}. Here, the extended, `quantum space-time'
has a single asymptotic region in the future, i.e., there are no
`baby universes'. In four dimensions, this is an
\emph{assumption}. It is motivated by two considerations: i) the
situation in the CGHS model where detailed calculations are
possible and show that the quantum space-time has this property;
and ii) experience with the action of the Hamiltonian constraint
in the spherically symmetric midi-superspace in four dimensions.
However, only detailed calculations can decide whether this
assumption is borne out. Since our goal in this paper is only to
point out the existence of a possible space-time description in
which information can be recovered at future null infinity, for
our purposes it suffices to note only that none of the existing
arguments rule out this mechanism.

We will refer to figure \ref{NonSingH} as a `Penrose diagram'
where the inverted commas will serve as a reminder that we are not
dealing with a purely classical space-time. Throughout the quantum
evolution, the pure state remains pure and so we again have a pure
state on ${\scri^+}$. In this sense there is no information loss.
Noteworthy features of this `Penrose diagram' are the following.

i) \textsl{Effect of the resolution of the classical singularity:}
Region marked I is well-approximated by a classical geometry.
Modulo small quantum fluctuations, this geometry is determined via
Einstein's equations by the classical data on ${\scri^-}$ at which
the incoming quantum state is peaked. The key difference between
figures \ref{Traditional} and \ref{NonSingH} is that while
space-time `ends' at the singularity in figure \ref{Traditional}
it does not end in figure \ref{NonSingH}. But there may not be
even an approximate classical space-time in the shaded region
representing the `deep Planck regime'.

ii) \textsl{Event horizon:} Since the shaded region does not have
a classical metric, it is not meaningful to ask questions about
causal relations between this region and the rest. Therefore,
although it {is} meaningful to analyze the causal structure (to an
excellent approximation) within each local semi-classical region,
due care must be exercised to address \textit{global} issues which
require knowledge of the metric on the entire space-time. This is
in particular the case for the notion of the event horizon, the
future boundary of the causal past of ${\scri}^+$. Because there
is no classical metric in the shaded region, while one can
unambiguously find some space-time regions which are in the past
of ${\scri}^+$, we can not determine what the \emph{entire} past of
${\scri}^+$ is. If we simply cut out this region and look at the
remaining classical space-time, we will find that the past is not
all of this space-time. But this procedure can not be justified
especially for purposes of quantum dynamics.  Thus, because the
geometry in the deep Planck regime is genuinely quantum
mechanical, the global notion of an event horizon ceases to be
useful. It may well be that there is a well-defined, new notion of
quantum causality and using it one may be able to reanalyze this
issue. However, the standard classical notion of the event horizon
is `transcended' because of absence of a useful
classical metric in the deep Planck region.%
\footnote{Some authors \cite{s'thw,sh3} have suggested that there
may be a classical metric on entire space-time but Einstein's
classical equations would be violated in the deep Planck region,
resulting in a metric which is continuous (or better behaved)
everywhere. Should this turn out to be the case the event horizon
would not just be `transcended' but simply disappear. Trapping
horizon would still be well-defined.}

iii) \textsl{Dynamical horizon:} Nonetheless, we can trust
classical theory in region I and this region will admit marginally
trapped surfaces. It is reasonable to expect that a spherical
dynamical horizon will be formed. It will be space-like and its
area will grow during collapse. In the classical theory, the
dynamical horizon will eventually settle down to a null, isolated
horizon which will coincide with (the late portion of) the event
horizon. However, in quantum theory eventually the horizon will
shrink because of Hawking radiation. While the black hole is
large, the process will be very slow. Semi-classical calculations
indicate that there is a positive flux of energy out of the black
hole. The dynamical horizon $H$ will now `evolve' into a time-like
membrane and its area loss will be dictated by the balance law
%
\be \label{2}\frac{dR}{dt} = - 8\pi G\, R^2\, T_{ab} \ell^a
\hat{r}^b \ee
where $R$ is the area radius of cross-sections of marginally
trapped 2-spheres in $H$, $\ell^a$ the (future directed) null
normal with vanishing expansion, and $\hat{r}^a$ is the unit
(outward) radial normal to $H$. (See Appendix B of the second
paper in \cite{ak2}). This process is depicted in figure
\ref{NonSingH}. The union of the dynamical horizon, the isolated
horizon and the timelike membrane constitutes the trapping
horizon. Thus, although we no longer have a well-defined notion
of an event horizon, we can still meaningfully discuss formation
and evaporation of the black hole using trapping horizons
because most of this process occurs in the semi-classical region
and, more importantly, because \emph{the notion of a trapping
horizon is quasi-local}. When the black hole is large, the
evaporation process is extremely slow. Therefore, it seems
reasonable to assume that the intuition developed from the quantum
geometry of isolated horizons \cite{abck,abk} will continue to be
valid. If so, the quantum geometry of the trapping horizon will be
described by the $U(1)$ Chern-Simons theory on a punctured $S^2$,
where the punctures result because the polymer excitations of the
bulk geometry pierce the dynamical horizon, endowing it with
certain area quanta. During the evaporation process, the punctures
slowly disappear, the horizon shrinks and quanta of area are
converted into quanta of the scalar field, seen as Hawking
radiation at infinity.%
\footnote{Equation (\ref{2}) relates the change in the area of the
time-like membrane part of the trapping horizon with the flux of
the energy flowing out of it. However, because of the dynamical
nature of geometry, there is no simple relation between this
ingoing flux at the time-like membrane and the energy carried by
the outgoing quanta on ${\scri}^+$. Indeed, not only are they
evaluated at very different locations, the two fluxes refer to
\emph{distinct} components of the stress-energy tensor,
$T_{ab}\ell^a \hat{r}^b$ at the horizon and $T_{ab}n^an^b$ at
${\scri}^+$.}
The existence, in the classical theory, of a meaningful
generalization of the first law of black hole mechanics to
dynamical horizons \cite{ak1,ak2} supports the view that the
process can be interpreted as evaporation of the dynamical
horizon.

\begin{figure}
\begin{center}
\includegraphics[height=3in,keepaspectratio]{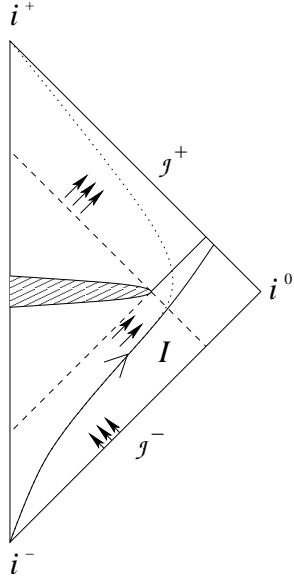}
\caption{The solid line with an arrow represents the world-line of
an observer restricted to lie in region I. While these observers
must eventually accelerate to reach ${\scri}^+$, if they are
sufficiently far away, they can move along an asymptotic time
translation for a long time. The dotted continuation of the world
line represents an observer who is not restricted to lie in region
I. These observers can follow an asymptotic time translation all
the way to $i^+$.} \label{NonSingAB}
\end{center}
\end{figure}
\begin{figure}
\begin{center}
\includegraphics[height=3.5in]{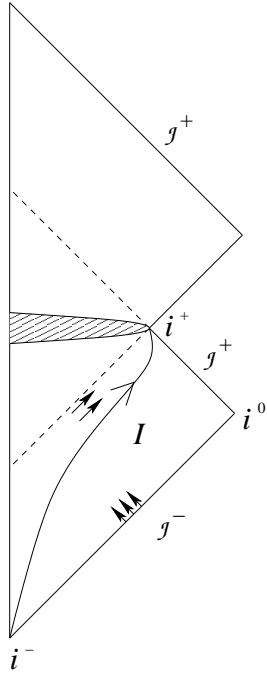}
\caption{The `would be' space-time if the black hole were to take
an infinite time to evaporate.} \label{Infinite}
\end{center}
\end{figure}


iv) \textsl{Reconciliation with the semi-classical information
loss:} Consider observers restricted to lie in region I (see
figure \ref{NonSingAB}). For a macroscopic black hole this
semi-classical region is very large. These observers would see the
radiation resulting from the evaporation of the horizon. This
would be approximately thermal, only approximately because, among
other things,  the space-time geometry is not fixed as in
Hawking's original calculation \cite{swh1}, but evolves slowly.
Although the full quantum state is `pure', there is no
contradiction because these observers look at only part I of the
system and trace over the rest which includes a purely quantum
part. In effect, for them space-time has a future boundary where
information is lost. Since the black hole is assumed to be
initially large, the evaporation time is long (about $10^{70}$
years for a solar mass black hole). Suppose we were to work with
an approximation that the black hole takes \emph{infinite} time to
evaporate. Then, the space-time diagram will be figure
\ref{Infinite} because the horizon area would shrink to zero only
at $i^+$. In this case, there would be an event horizon and
information would be genuinely lost for any observer in the
initial space-time; it would go to a second asymptotic region
which is inaccessible to observers in the initial space-time. Of
course this does not happen because the black hole evaporates only
in a finite time.

v) \textsl{`Recovery' of the `apparently lost' information:} Since
the black hole evaporates only in a finite amount of time, the
point at which the black hole shrinks to zero (or Planck) size
is \emph{not} $i^+$ and the space-time diagram looks like figure
\ref{NonSingAB} rather than figure \ref{Infinite}. Now, $i^+$ lies
to the `future' of the `deep Planck' region and there are
observers lying entirely in the asymptotic region going from $i^-$
to $i^+$ (represented by the dotted continuation of the solid line
in figure \ref{NonSingAB}). This family of observers will recover
the apparently lost correlations. Note that these observers always
remain in the asymptotic region where there is a classical metric
to an excellent approximation; they {never} go near the deep
Planck region. The total quantum state on ${\scri}^+$ will be pure
and will have the complete information about the initial state on
${\scri^-}$. It looked approximately thermal at early times, i.e.,
to observers represented by the solid line, only because they
ignore a part of space-time. The situation has some similarity
with the EPR experiment in which the two subsystems are first
widely separated and then brought together (see also \cite{fw}).

vi) \textsl{Entropy:} Since the true state is always pure, one
might wonder what happens to black hole entropy. It is only the
observers in region I that `sense' the presence of a black hole.
In the quantum geometry approach to black hole entropy, entropy is
not an absolute concept associated objectively with a space-time.
Rather, it is associated with a family of observers who have
access to only a part of space-time.  Indeed, the entropy of an
isolated horizon calculated in \cite{abk} referred to the family
of observers for whom the isolated horizon serves as the internal
boundary of accessible space-time. So, for observers restricted to
region I, that entropy calculation is still meaningful, at least
so long as the black hole is macroscopic (i.e., the area of
marginally trapped surfaces on $H$ is much larger than Planck
area). And it is these observers who see the (approximate) Hawking
radiation. More precisely, since these observers have access only
to observables of the type $A_{\rm I}\otimes 1$, they trace over
the part of the system not in I, getting a density matrix
$\rho_{\rm I}$ on the Hilbert space ${\cal H}_{\rm I}$. Entropy
for them is simply $Tr_{\rm I} \rho_{\rm I} \ln \rho_{\rm I}$. Had
there been a true singularity `ending' the space-time, this
entropy would have become objective in the sense that it would be
associated with \emph{all} observers who do not fall into the
singularity.

\section{Concluding Remarks}
\label{s4}

In the last two sections we used a quantum gravity perspective to
argue that information loss is not inevitable in the space-time
description of black hole evaporation. The qualitative difference
between figures \ref{Traditional} and \ref{NonSingH} arises
essentially from the fact that the singularity is resolved in
quantum geometry, as per a general expectation that a satisfactory
quantum theory of gravity should not have infinities. In this
sense the paradigm shift is well-motivated. Furthermore,
conclusions of the traditional paradigm drawn from the usual
space-time diagram \ref{Traditional} are not simply discarded. For
a large black hole, they continue to be approximately valid for a
very long time. Figure \ref{NonSingAB} clarifies the approximation
involved. However, from the \emph{conceptual perspective of
fundamental physics}, conclusions drawn from the \emph{complete}
space-time diagram \ref{NonSingH} are qualitatively different from
the standard ones. A pure state from ${\scri}^-$ evolves to a pure
state on ${\scri}^+$ and there is no obstruction in quantum theory
to evolving the final state on ${\scri}^+$ backwards to recover
full space-time. However it is likely that the resulting geometry
would fail to be globally classical. In the shaded region, it is
likely to be genuinely quantum mechanical, described only in terms
of the quantum geometry states (i.e., in terms of spin-networks).
In the region in which one can introduce classical geometry to
an excellent approximation, it is meaningful to speak of
marginally trapped surfaces, trapping horizons and null infinity
${\scri}^\pm$.%
\footnote{Because of the presence of the purely quantum part, the
space-time is not asymptotically simple \cite{rp}; the classical
region admits null geodesics which do not end on ${\scri}^\pm$.
However, it is asymptotically flat and admits a global null
infinity in the sense of \cite{ax}.}
What `evaporates' is the area of the trapping horizon.

From the perspective of this paradigm, the conclusion that a pure
state must evolve to a mixed state results if one takes the
classical space-time diagram \ref{Traditional}, \emph{including
the singular boundary in the future}, too seriously.%
\footnote{Perhaps an analogy from atomic physics would be to base
the analysis of the ground state of the hydrogen atom on the zero
angular momentum, classical electron trajectories, all of which
pass through the `singularity' at the origin.}
In the cosmological context, a combination of detailed analytical
and numerical calculations \cite{aps} has recently shown that
quantum geometry is well-defined at the classical big-bang
singularity; backward quantum evolution enables one to pass
through it; and, on the other side, there is again a classical
space-time. Thus, quantum geometry in the deep Planck regime
serves to bridge two large classical regimes. Classical
singularity is only a reflection of the failure of the mean field
approximation and quantum geometry is defined on a larger
manifold. Our paradigm is based on the assumption that the
situation is qualitatively similar with black hole
singularities. If this assumption is borne out, pure states will
evolve to pure states, without any information loss provided the
analysis pays due respect to this space-time extension.

The two dimensional analog of our paradigm is realized quite well
by CGHS black holes. However, 2 dimensional models have special
features that are not shared by higher dimensional theories. To
carry out the analogous analysis in 4 dimensions, one would have to
complete several difficult steps: \\
i) Discussion of quantum dynamics in the spherically symmetric
midi-superspace \cite{bs}. To be directly useful, we would need to
introduce a satisfactory generalization of the notion of `time'
used in \cite{ab1,aps}; \\
ii) demonstration of the semi-classical behavior of the quantum
state in regions where the dynamical horizon grows and the
time-like membrane shrinks (in the regime where its area is
large); \\
iii) extension of the available theory \cite{abk} of quantum
geometry from isolated to slowly evolving dynamical horizons;
and\\
iv) establishing that the quantum state becomes semi-classical
again on the `other side' of what was a classical singularity,
with a single asymptotic region. \\
Note, however, that any approach to quantum gravity will have
to resolve similar issues if it is to provide a detailed
`space-time description' of the black hole evaporation in the
Lorentzian framework. In particular, all discussions beyond the
semi-classical approximation that we are aware of implicitly
assume that there is a classical space-time in the future.

Finally, in this paradigm correlations are restored by part of the
state that passes through the singularity and emerges on
${\scri}^+$ to the future of region I of figure \ref{NonSingH}.
Therefore, it is presumably necessary that this part should carry
a non-trivial fraction of the total ADM mass of space-time (see,
however, \cite{fw}). This seems physically plausible because one
expects non-trivial space-time curvature also on the `other side
of the singularity'. However, whether this is realized in detailed
calculations remains to be seen. Thus, the paradigm is based on
pieces of calculations and analogy to the CGHS model, rather than
a systematic detailed analysis. Recall, however, that the
traditional reasoning that led to figure \ref{Traditional} was
based on general considerations and plausibility arguments and a
systematic analysis of the viability of approximations is still
not available. Nonetheless, it led to a paradigm which proved to
be valuable in focussing discussions. Our hope is that that the
paradigm presented here will play a similar role.

\emph{Remark:} After this work was posted on the archives, we
became aware of two discussions of black hole evaporation which
feature space-time diagrams similar to figure \ref{NonSingH}. The
first is due to Stephens, 't Hooft and Whiting \cite{s'thw} which
appeared more than a decade ago and the second is due to Hayward
\cite{sh3} which appeared very recently. In the first, one draws a
distinction between hard matter which creates curvature and soft
matter whose effect on gravity is negligible. A detailed
calculation is carried out in a 2-dimensional model, where the
focus is on the soft matter. The main idea is to first assume that
quantum gravity effects would halt the collapse and cause a bounce
and then do a calculation analogous to that of Hawking's
\cite{swh1} on this modified but classical background geometry.
The result is that although pure states evolve to pure states, in
the appropriate portion of ${\scri}^+$, the state is approximately
thermal. This scenario is similar to ours in that the space-time
under consideration has no singularity; pure states evolve to pure
states; and expectations based on semi-classical considerations
are not just discarded but recovered in a precise sense. However,
there are also some important differences. If our paradigm is
realized by detailed calculations, all matter would be `hard';
singularity would be resolved by specific quantum gravity effects;
and a genuinely quantum mechanical geometry would bridge the
space-time of classical general relativity with a new classical
space-time. In contrast to \cite{s'thw}, the new portion in the
geometry of any one space-time will not correspond to a simple
time-reversal of the standard, collapsing portion. Hayward's
considerations \cite{sh3} are different from those of
\cite{s'thw}. As in the current paradigm, he emphasizes trapping
horizons and his space-time diagram is closer to ours, especially
for a massless Klein-Gordon source. In particular, his space-time
is a singularity-free extension of standard one and the collapsing
matter re-emerges on ${\scri}^+$, in addition to the Hawking
radiation. However, he assumes that space-time will have a $C^2$
metric everywhere (which, however, violates the classical field
equations near what was the singularity), and the collapsing
matter which re-emerges is treated classically. Apart from the
Hawking radiation, genuine quantum considerations do not appear to
play a significant role. Recent numerical evolutions in quantum
cosmology \cite{aps} indicate that there may well exist initial
states on ${\scri}^-$ for which the physics of our deep Planck
regime can be approximated by an effective continuum classical
geometry. If this does happen for black hole space-times, then our
paradigm would essentially reduce to Hayward's in those
situations.

\section*{Acknowledgements:}

We thank Alex Corichi, Gary Horowitz, Bei-Lok Hu, Ted Jacobson,
Daniel Sudarsky and especially Jim Hartle, Sean Hayward and Don
Marolf for stimulating discussions. This work was supported in
part by NSF grants PHY-0090091, and PHY-0354932, the Alexander von
Humboldt Foundation, the C.V. Raman Chair of the Indian Academy of
Sciences and the Eberly research funds of Penn State.

\end{document}